\documentclass[parskip=full,12pt]{article}

\usepackage{soul}
\usepackage{amssymb}
\usepackage{amsthm}
\usepackage{amsmath}
\usepackage{booktabs}
\usepackage{graphicx}
\usepackage{natbib}
\usepackage{url}
\usepackage{color,soul}
\usepackage{multirow}
\usepackage[top=1in, bottom=1.3in, left=0.875in, right=0.875in]{geometry}
\usepackage[linesnumbered,ruled]{algorithm2e}
\usepackage{tikz}
\usetikzlibrary{er,positioning,arrows.meta,calc}
\usepackage[labelformat=simple]{subcaption}

\usetikzlibrary{er,positioning,arrows.meta,arrows,shapes,calc}

\tikzstyle{block} = [rectangle, draw, fill=white, 
    text width=7cm, text centered, rounded corners, minimum height=3em]

\tikzset{
     arrow/.style = { thick,  ->, >=Triangle},
}

\def\spacingset#1{\renewcommand{\baselinestretch}%
{#1}\small\normalsize} \spacingset{1}

\newtheorem{lemma}{Lemma}[section]
\def\spacingset#1{\renewcommand{\baselinestretch}%
{#1}\small\normalsize} \spacingset{1}

\title{\textbf{Hybrid Tree-based Models for Insurance Claims}}

\author{Zhiyu Quan\thanks{Department of Mathematics, University of Connecticut, 341 Mansfield Road, Storrs, CT, 06269-1009, USA. Email: \texttt{zhiyu.quan@uconn.edu}.} \and Zhiguo Wang\thanks{Department of Mathematics, University of Connecticut, 341 Mansfield Road, Storrs, CT, 06269-1009, USA. Email: \texttt{zhiguo.wang@uconn.edu}.} \and Guojun Gan\thanks{Department of Mathematics, University of Connecticut, 341 Mansfield Road, Storrs, CT, 06269-1009, USA. Email: \texttt{guojun.gan@uconn.edu}.} \and Emiliano A. Valdez\thanks{Department of Mathematics, University of Connecticut, 341 Mansfield Road, Storrs, CT, 06269-1009, USA. Email: \texttt{emiliano.valdez@uconn.edu}.}}

\begin{document}
\maketitle

\begin{abstract}

Two-part models and Tweedie generalized linear models (GLMs)  have been used to model loss costs for short-term insurance contract. For most portfolios of insurance claims, there is typically a large proportion of zero claims that leads to imbalances  resulting in inferior prediction accuracy of these traditional approaches. This article proposes the use of tree-based models with a hybrid structure that involves a two-step algorithm as an alternative approach to these traditional models. The first step is the construction of a classification tree to build the probability model for frequency. In the second step, we employ elastic net regression models at each terminal node from the classification tree to build the distribution model for severity. This hybrid structure captures the benefits of tuning hyperparameters at each step of the algorithm; this allows for improved prediction accuracy and tuning can be performed to meet specific business objectives. We examine and compare the predictive performance of such a hybrid tree-based structure in relation to the traditional Tweedie model using both real and synthetic datasets. Our empirical results show that these hybrid tree-based models produce more accurate predictions without the loss of intuitive interpretation.

\vspace{0.75cm}

\noindent \textbf{Keywords:} pure premium; Tweedie GLM; classification trees; regularized regression; parameter tuning.

\end{abstract}

\newpage

\section{Prelude} \label{sec:intro}

Building regression models for insurance claims presents several challenges. The process can be particularly difficult for individual insurance policies where a large proportion of the claims are zeros and for those policies with a claim, the losses typically exhibit skewness. The degree of skewness in the positive loss distribution varies widely among different lines of insurance business. For example, within a class of automobile insurance policies, the losses arising from damages to property may be medium-tailed but those arising from liability-related claims may be long-tailed. For short-term insurance contracts, the number of claims is referred to as the claim frequency while the amount of claims, conditional on the occurrence of a claim, is the claim severity. The addition of covariates in a regression context within each component allows for capturing the heterogeneity of the individual policies. These two components are used as predictive models for pure premiums and for aggregate or total claims in a portfolio.

There is a long history of studying insurance claims based on the two-part framework. Several well-known families of distributions for claim frequencies and claim severities are well documented in \cite{klugman2012}. As more observed data become increasingly available, additional complexities to traditional regression linear models have been introduced both in the literature and in practice; for example, the use of zero-truncated or hurdle models has been discussed to accommodate some of the shortcomings of traditional models. Interestingly, the Tweedie regression models for insurance claims have been particularly popular because it is adaptable to a mixture of zeros and non-negative insurance claims.  \cite{smyth2002fitting} described the Tweedie compound Poisson regression models and calibrated such models using a Swedish third party insurance claims dataset; the models were fitted to examine dispersion within the GLM framework. The Tweedie distribution can be treated as a reparameterization of the compound Poisson-gamma distribution, so that model calibration can be done within the GLM framework. \citet{xacur2015generalised} compared these two approaches and concluded that there is no clear superior method; the paper also described the advantages and disadvantages of the two methods. Tweedie GLM presents a larger pure premium and implies both a larger claim frequency and claim severity due to the constant scale (dispersion) parameter. In other words, the mean increases with variance. The constant scale parameter also forces an artificial relationship between the claim frequency and the claim severity. Tweedie GLM does not have an optimal coefficient, and it leads to a loss of information because it ignores the number of claims. On the other hand, Tweedie GLM has fewer parameters to estimate and is thus more parsimonious than the two-part framework. When the insurance claims data presents small losses due to low frequencies, the two-part framework most likely overlooks the internal connection between the low frequency and the subsequent small loss amount. For example, the frequency model often indicates zero number of claims for small claim policies, which leads to a zero loss prediction. Additional works related to Tweedie regression can be found in \cite{frees2014predictive} and \cite{jorgensen1987exponential}. 

Within the GLM framework, the claim frequency component is unable to accurately accommodate imbalances caused by the large proportion of zero claims. Simultaneously, a significant limitation is that the regression structure of the logarithmic mean is restricted to a linear form, which may be too inflexible for real applications. With rapid expansion of available data for improved decision making, there is a growing appetite in the insurance industry for expanding its toolkit for data analysis and modeling. However, the industry is unarguably highly regulated so that there is pressure for actuaries and data scientists to provide adequate transparency in modeling techniques and results. To find a balance between modeling flexibility and interpretation, in this paper, we propose a nonparametric model using tree-based models with a hybrid structure. Since \citet{breiman1984} introduced the Classification and Regression Tree (CART) algorithm, tree-based models have gained momentum as a powerful machine learning approach for decision making in several disciplines. The CART algorithm involves separating the explanatory variable space into several mutually exclusive regions that, as a result, creates a nested hierarchy of branches resembling a tree structure. Each separation or branch is referred to as a node. Each of the bottom nodes of the decision tree, called terminal nodes, has a unique path for observable data to enter the region. Once the decision tree is constructed, it is possible to use paths to locate the region or terminal node to which a new set of explanatory variables will belong.

Prior to further exploring the hybrid structure, it is worth mentioning that a similar structure called Model trees, with an M5 algorithm, was first described in \cite{quinlan1992learning}. In the M5 algorithm, it constructs tree-based piecewise linear models. Regression trees assign a constant value to the terminal node as the fitted value. However, Model trees use a linear regression model at each terminal node to predict the fitted value for observations that reach that terminal node. Regression trees are a special case of Model trees. Both regression trees and Model trees employ recursive binary splitting to build a fully grown tree. Thereafter, both algorithms use a cost-complexity pruning procedure to trim the fully grown tree back from each terminal node. The primary difference between regression trees and Model trees algorithms is that for the latter step, each terminal node is replaced by a regression model instead of a constant value. The explanatory variables that serve to build that regression models are generally those that participate in the splits within the subtree that will be pruned. In other words, explanatory variables in each node are located beneath the current terminal node in the fully grown tree.

It has been demonstrated that Model trees have advantages over regression trees in both model simplicity and prediction accuracy. Model trees produce decision trees that are considered not only relatively simple to understand, but are also efficient and robust. Additionally, they are able to exploit local linearity within the dataset. When prediction performance are compared, it is worth mentioning the difference in the range of predictions produced between the traditional regression trees and these Model trees. Accordingly, regression trees are only able to give a prediction within the range of observed values in the training dataset. However, Model trees are able to extrapolate prediction range because of the use of the regression models at the terminal node. For further discussion of Model trees and M5 algorithm, please see \cite{quinlan1992}.

Inspired by the structure and advantages of Model trees when compared to traditional regression trees, we develop a similar algorithm that can be uniquely applied to insurance claims because of its two-part nature. In this paper, we present the hybrid tree-based models as a two-step algorithm: the first step builds a classification tree to identify membership of claim occurrence and the second step uses a penalized regression technique to determine the size of the claim at each of the terminal nodes, taking into account the available explanatory variables. In essence, hybrid tree-based models for insurance claims as described in this paper integrate the peculiarities of both the classification tree and the linear regression in the modeling process. These sets of models are suitably described as hybrid structures.

We have organized the rest of this paper as follows. In Section \ref{sec:HT}, we provide technical details of the algorithm arising from hybrid tree-based models applicable to insurance claims, separating the modeling of claim frequency and claim severity. In section \ref{sec:sim}, we create a synthetic dataset based on a simulation produced from a true Tweedie regression model to investigate the predictive accuracy of our hybrid tree-based models when compared to the Tweedie GLM.  We introduced some noise to the simulated data in order to be able to make a reasonable comparison. In section \ref{sec:APP}, using empirical data drawn from a portfolio of general insurance policies, we present the estimation results and compared the prediction accuracy based on a hold-out sample for validation. Section \ref{sec:visual} provides visualization of the results from hybrid-tree models to allow for better interpretation. We conclude in Section \ref{sec:conclude}. 

\section{Description of hybrid tree-based models} \label{sec:HT}

Hybrid tree-based models utilize the information arising from both the frequency and the severity components. As already alluded in Section \ref{sec:intro}, it is a two-stage procedure. In the first stage, we construct a classification tree-based model for frequency. In the subsequent stage, we employ a penalized linear regression model at each terminal node within the tree-based model, based on the severity information. Such models can be drawn from the hybrid structure, which is an ecosystem, to accommodate the variety of the dataset. This hybrid structure captures all the advantageous features of tree-based models and the penalized regression models.

We can determine the type of classification trees that can be adapted in the frequency according to the information drawn from the insurance dataset. If the dataset records only whether claims were reported or not, we can construct a classification tree for a binary response variable. If the dataset records additional information of a number of claims, we can construct trees for a count response variable. For the binary classification, we can employ some of the most popular CART algorithm, to list a few: (a) efficient algorithm C4.5 \citep{quinlan1992}, with a current updated version C5.0, (b) unbiased approaches like Generalized, Unbiased, Interaction Detection, and Estimation (GUIDE) \citep{loh2009improving}, or (c) Conditional Inference Trees (CTREE) \citep{hothorn2006unbiased}. For the count response variable, we can apply, to list a few: (a) piecewise linear Poisson using the CART algorithm called SUPPORT \citep{chaudhuri1995generalized}, (b) Poisson regression using GUIDE \citep{loh2006regression}, or (c) MOdel-Based recursive partitioning (MOB) \citep{zeileis2008model}.

After the completion of the classification tree structure, we then employ a linear regression model to each of the terminal nodes. The simplest model includes GLM \citep{nelder1972glm} with different families and regularized regression like elastic net regularization \citep{zou2005reg} amongst others.

Hybrid tree-based models build an ecosystem that utilizes modern techniques from both traditional statistics and machine learning. In the subsequent subsections, for simplicity, we only show a simple hybrid tree-based model without exploring all of the possible combinations of different algorithms suitable at each stage of the procedure.

\subsection{Claim frequency} \label{frequency}

To illustrate the use of hybrid tree-based models, we select the well-known CART algorithm for binary classification and least squares regression with elastic net regularization. Elastic net regularization can perform variable selection and improve the prediction accuracy when compared to traditional linear regressions without penalty. In order to be self-contained, we illustrate this simple hybrid tree-based model with sufficient details. See also \cite{james2013}.

We denote the response variable as \(\textbf{Y}\), the sample space as \(\mathcal{Y}\), and \(n\) as the number of observations. The \(i\)th sample with \(p\)-dimensional explanatory variables is denoted as \(\textbf{x}_i = (x_{i1},x_{i2}\ldots,x_{ip})\) \(i=1 \dots n\), which is sampled from the space \(\mathcal{X} = \mathcal{X}_{1} \times \ldots \times \mathcal{X}_{p}\). For example, we can separate each claim \(y_i\) into \(y_{i_{f}}\), the claim occurrence or the number of claims, and \(y_{i_{s}}\), the claim severity.

In the CART algorithm, a binary classification tree, denoted by \(T(\textbf{x},\Theta)\), is produced by partitioning the space of the explanatory variables into \(M\) disjoint regions \(R_{1},R_{2}, \ldots, R_{M}\) and then assigning a boolean \(\beta_{f_{m}}\) for each region \(R_m\), for \(m=1,2, \ldots, M\). Given a classification tree, each observation can then be classified based on the expression
\begin{equation} \label{eq:1}
T(\textbf{x},\Theta) = \sum_{m=1}^{M} \beta_{f_{m}} \mathbf {1}_{R_{m}}(\textbf{x}_i), 
\end{equation}
where \(\Theta = \{R_{m},\beta_{f_{m}}\}_{m=1}^{M}\) denotes the partition with the assigned boolean. To be more specific, boolean \(\beta_{f_{m}} = 1\) when the majority of observations in region \(R_{m}\) have a claim; otherwise it is zero.

The traditional classification loss functions used in the classification tree, in Equation (\ref{eq:1}), are described as follows:
\begin{itemize}
\item Misclassification error: $\quad \text{Mis}(p) = 1-\max(p,1-p)$
\smallskip
\item Gini index: $\quad \text{Gini}(p) = 2p(1-p)$
\smallskip
\item Cross-entropy or deviance: $\quad \text{Entropy}(p) =-p \log p - (1-p) \log (1-p)$
\end{itemize}
where \(p\) is proportion of one class in the node. For multi-class loss functions, see \cite{hastie2009}.

The default CART algorithm uses Gini index as a loss function. The regions in the classification tree are determined according to recursive binary splitting. First in the process of splitting is the discovery of one explanatory variable \(x_{\cdot j}\) which best divides the data into two subregions; for example, these regions are the left node \(R_L(j,s)=\{\textbf{x}_{i}|x_{\cdot j} < s \}\) and the right node \(R_R(j,s)=\{\textbf{x}_{i}|x_{\cdot j} \ge s \}\) in the case of a continuous explanatory variable. This division is determined as the solution to
\begin{equation}  \label{eq:2}
\underset{j,s}{\mathrm{argmin}} {\quad w_L\,\text{Gini}(p_L) + w_R\,\text{Gini}(p_R)}, \quad 
\text{for any} \ j \ \text{and} \ s,
\end{equation}
where \(w_.\) is the weight for the subregion determined by the number of observations split into subregion divided by the total number of observations before the split, and \(p_.\) is the proportion of one class in the subregion. Subsequently, the algorithm looks for the next explanatory variable with the best division into two subregions and this process is applied recursively until meeting some predefined threshold or reaching a minimum size of observations in the terminal node.

To control for model complexity, we can use \emph{cost-complexity pruning} to trim the fully grown tree \(T_0\). We define the loss in region \(R_m\) by
\[
L_m(T) = w_m\,\text{Mis}(p_m).
\]
For any subtree \(T \subset T_0\), we denote the number of terminal nodes in this subtree by \(|T|\). To control the number of terminal nodes, we introduce the tuning hyperparameter \(\alpha \geq 0\) to the loss function by defining the new cost function as
\begin{equation} \label{eq:3}
C_\alpha(T) = \sum_{m=1}^{|T|}L_m(T) + \alpha|T|.
\end{equation}
Clearly, according to this cost function, the tuning hyperparameter penalizes large numbers of terminal nodes. The idea then is to find the subtrees \(T_\alpha \subset T_0\) for each \(\alpha\), and choose the subtree that minimizes \(C_\alpha(T)\) in Equation (\ref{eq:3}). Furthermore, the tuning hyperparameter \(\alpha\) governs the tradeoff between the size of the tree (model complexity) and its goodness of fit to the data. Large values of \(\alpha\) result in smaller trees (simple model) and as the notation suggests, \(\alpha = 0\) leads to the fully grown tree \(T_0\). Additional tuning is done to control the tree depth through \(maxdepth\), which is the maximum depth of any node of the final tree.

\subsection{Claim severity} \label{severity}

After building the classification tree structure, we next apply a linear regression on the terminal nodes to model severity. In controlling for model complexity, we set a threshold to determine if we should build a linear regression or directly assign zero for the terminal. For example, if \(zeroThreshold\) is 80\%, then we should directly assign zero to the terminal nodes that contain more than 80\% of zero claims. Furthermore, if the terminal node contains less than a certain number of observations, say 40, we can directly use the mean as the prediction similar to regression trees. Otherwise, we need to build a linear regression on the terminal nodes. While any member of the GLM exponential family that is suitable for continuous claims can be used, we find that the special case of GLM Gaussian, or just ordinary linear regression, is sufficiently suitable.

At each terminal node \(R_{m}\), the linear coefficient \(\mathbf{\boldsymbol{\beta}}_{s_{m}} = (\beta_{s_{m}0}, \beta_{s_{m}1}, \dots, \beta_{s_{m}p})^T\) can be determined by:
\begin{equation} \label{eq:4}
\mathbf{\boldsymbol{\widehat{\beta}}}_{s_{m}}  = 
\underset{\mathbf{\boldsymbol{\beta}_{s_{m}}}} {\operatorname{arg\,min}} \ \frac{1}{n} \sum_{i=1}^n \ell \Big(y_i, \beta_{s_{m}0} + \sum_{j=1}^p x_{ij} \beta_{s_{m}j}\Big)
\end{equation}
where \(\ell(y_i, \beta_{s_{m}0} + \sum_{j=1}^p x_{ij} \beta_{s_{m}j})\) is the negative log-likelihood for sample \(i\). For the Gaussian family, denoting the design matrix as $\textbf{X}$, the coefficient is well known as
\[
\mathbf{\boldsymbol{\widehat{\beta}}}_{s_{m}}  = (\textbf{X}^T\textbf{X})^{-1}\textbf{X}^T \textbf{Y}.
\]

Ridge regression \citep{hoerl1970ridge} achieves better prediction accuracy compared to ordinary least squares because of bias-variance trade-off. In other words, the reduction in the variance term of the coefficient is larger than the increase in its squared bias. It performs coefficient shrinkage and forces its correlated explanatory variables to have similar coefficients. In ridge regression, at each terminal node \(R_{m}\), the linear coefficient \(\mathbf{\boldsymbol{\beta}_{s_{m}}}\) can be determined by
\begin{equation} \label{eq:5}
\mathbf{\boldsymbol{\widehat{\beta}}}_{s_{m}}  = \underset{\mathbf{\boldsymbol{\beta}_{s_{m}}}}{\operatorname{arg\,min}} \ \frac{1}{n} \sum_{i=1}^n \ell \Big(y_i, \beta_{s_{m}0} + \sum_{j=1}^p x_{ij} \beta_{s_{m}j}\Big) + \lambda \sum_{j=1}^p \beta_{s_{m}j}^2
\end{equation}
where \(\lambda\) is a tuning hyperparameter that controls shrinkage and thus, the number of selected explanatory variables. For the following discussion, we assume the values of \(x_{ij}\) are standardized so that
\[
\frac{1}{n} \sum_{i=1}^n x_{ij} = 0 \quad \text{and} \quad \sum_{i=1}^n x_{ij}^2=1.
\]
If the explanatory variables do not have the same scale, the shrinkage may not be fair. In the case of ridge regression within the Gaussian family, the coefficient in Equation (\ref{eq:5}) can be shown to have the explicit form
\[
\mathbf{\boldsymbol{\widehat{\beta}}}_{s_{m}}  = (\textbf{X}^T\textbf{X} + \lambda \textbf{I})^{-1}\textbf{X}^T\textbf{Y},
\]
where $I$ is the identity matrix of appropriate dimension.

Ridge regression does not automatically select the important explanatory variables. However, LASSO regression \citep{tibshirani1996regression} has the effect of sparsity, which forces the coefficients of the least important explanatory variable to have a zero coefficient therefore making the regression model more parsimonious. In addition, LASSO performs coefficient shrinkage and selects only one explanatory variable from the group of correlated explanatory variables. In LASSO, at each terminal node \(R_{m}\), the linear coefficient \(\mathbf{\boldsymbol{\beta}}_{s_{m}}\) can be determined by
\begin{equation} \label{eq:6}
\mathbf{\boldsymbol{\widehat{\beta}}}_{s_{m}}  = \underset{\mathbf{\boldsymbol{\beta}_{s_{m}}}}{\operatorname{arg\,min}} \ \frac{1}{n} \sum_{i=1}^n \ell \Big(y_i, \beta_{s_{m}0} + \sum_{j=1}^p x_{ij} \beta_{s_{m}j}\Big) + \lambda \sum_{j=1}^p |\beta_{s_{m}j}|,
\end{equation}
where \(\lambda\) is a tuning hyperparameter that controls shrinkage. For regularized least squares with LASSO penalty, Equation (\ref{eq:6}) leads to the following quadratic programming problem
\[
\mathbf{\boldsymbol{\widehat{\beta}}}_{s_{m}}  = \underset{\mathbf{\boldsymbol{\beta}}_{s_{m}}}{\operatorname{arg\,min}} \ \frac{1}{2n} \sum_{i=1}^n \Big(y_i - \beta_{s_{m}0} - \sum_{j=1}^p x_{ij} \beta_{s_{m}j}\Big)^2 + \lambda \sum_{j=1}^p |\beta_{s_{m}j}|.
\]
Originally, \citet{tibshirani1996regression} used the combined quadratic programming method to numerically solve for \(\mathbf{\boldsymbol{\widehat{\beta}}}_{s_{m}}\). In a later development, \citet{fu1998penalized} proposed the shooting method and \citet{friedman2007pathwise} redefined shooting as the coordinate descent algorithm which is a popular algorithm used in optimization.

To illustrate the coordinate descent algorithm, we start with \(\mathbf{\boldsymbol{\widehat{\beta}}}_{s_{m}} = \textbf0\). Given \(\beta_{s_{m}j}, \ j \neq k,\) the optimal \(\beta_{s_{m}k}\) can be found by
\[
\widehat{\beta}_{s_{m}k}= \underset{\beta_{s_{m}k}}{\operatorname{arg\,min}} \ \frac{1}{2n} \sum_{i=1}^n \Big(y_i - \beta_{s_{m}0} - \sum_{j \neq k} x_{ij} \beta_{s_{m}j}- x_{ik}\beta_{s_{m}k}\Big)^2 + \lambda \sum_{j \neq k} |\beta_{s_{m}j}| + \lambda |\beta_{s_{m}k}|,
\]
where \(\lambda \sum_{j \neq k} |\beta_{s_{m}j}|\) is constant and can be dropped. We denote \(\tilde{y_i} = y_i - \beta_{s_{m}0} - \sum_{j \neq k} x_{ij} \beta_{s_{m}j}\). It follows that
\[
\begin{aligned}
\widehat{\beta}_{s_{m}k}&= \underset{\beta_{s_{m}k}}{\operatorname{arg\,min}} \ \frac{1}{2n} \sum_{i=1}^n (\tilde{y_i}- x_{ik}\beta_{s_{m}k})^2 +  \lambda |\beta_{s_{m}k}| \\
&= \underset{\beta_{s_{m}k}}{\operatorname{arg\,min}} \ \frac{1}{2n} 
\Big( \sum_{i=1}^n \tilde{y_i}^2 + \beta_{s_{m}k}^2 \sum_{i=1}^n x_{ik}^2 - 2\beta_{s_{m}k} \sum_{i=1}^n\tilde{y_i}x_{ik} \Big) + \lambda |\beta_{s_{m}k}| \\
&= \underset{\beta_{s_{m}k}}{\operatorname{arg\,min}} \ \frac{1}{2n} 
\Big(\sum_{i=1}^n x_{ik}^2\Big) \Big(\beta_{s_{m}k} - \frac{\sum_{i=1}^n\tilde{y_i}x_{ik}}{\sum_{i=1}^n x_{ik}^2} \Big)^2 + \lambda |\beta_{s_{m}k}| + \text{constant}  \\
&= \underset{\beta_{s_{m}k}}{\operatorname{arg\,min}} \ 
\Big(\beta_{s_{m}k} - \frac{\sum_{i=1}^n\tilde{y_i}x_{ik}}{\sum_{i=1}^n x_{ik}^2} \Big)^2 +
\frac{\lambda}{\frac{1}{2n}\sum_{i=1}^n x_{ik}^2} |\beta_{s_{m}k}|.
\end{aligned}
\]
This can be solved by the Soft-thresholding Lemma discussed in \citep{donoho1995adapting}.

\begin{lemma} \label{lem:st}
\textbf{(Soft-thresholding Lemma)} The following optimization problem
\[
\widehat\beta(t,\lambda) = \underset{\beta }{\operatorname{arg\,min}} \ (\beta-t)^2 + \lambda |\beta|
\]
has the solution of
\[
\begin{aligned}
\widehat\beta(t,\lambda) &= (|t|-\lambda/2)^+ \text{sgn}(t) \\
&= \begin{cases}
  t-\frac{\lambda}{2}, & \text{if } t>0 \text{ and } \frac{\lambda}{2} < |t|, \\
  t+\frac{\lambda}{2}, & \text{if } t<0 \text{ and } \frac{\lambda}{2} < |t|, \\
  0, & \frac{\lambda}{2} \ge |t|. \\
\end{cases}
\end{aligned}
\]
\end{lemma}

Therefore, \(\widehat{\beta}_{s_{m}k}\) can be expressed as
\[
\widehat{\beta}_{s_{m}k} = \left(\frac{\sum_{i=1}^n |\tilde{y_i}x_{ik}| }{\sum_{i=1}^n x_{ik}^2} - \frac{\lambda}{\frac{1}{n}\sum_{i=1}^n x_{ik}^2}\right)^+  \, \text{sgn} \Big( \sum_{i=1}^n \tilde{y_i}x_{ik} \Big)
\]
For \(j = 1, \dots , p\), update \(\widehat{\beta}_{s_{m}k}\) by soft-thresholding when \(\beta_{s_{m}j}, j\neq k\) takes the previously estimated value. Repeat loop until convergence.

\citet{zou2005reg} pointed out that LASSO has few drawbacks. For example, when the number of explanatory variables \(p\) is larger than the number of observations \(n\), LASSO selects at most \(n\) explanatory variables. Additionally, if there is a group of pairwise correlated explanatory variables, LASSO randomly selects only one explanatory variable from this group and ignores the rest. \citet{zou2005reg} also empirically showed that LASSO has subprime prediction performance compared to ridge regression. Therefore, elastic net regularization is proposed which uses a convex combination of ridge and LASSO regression. Elastic net regularization is able to better handle correlated explanatory variables and perform variable selection and coefficient shrinkage. In elastic net regularization, at each terminal node \(R_{m}\), the linear coefficient \(\mathbf{\boldsymbol{\beta}}_{s_{m}}\) can be determined by
\begin{equation} \label{eq:7}
\mathbf{\boldsymbol{\widehat{\beta}}}_{s_{m}}  = \underset{\mathbf{\boldsymbol{\beta}}_{s_{m}}}{\operatorname{arg\,min}} \ \frac{1}{n} \sum_{i=1}^n \ell\Big(y_i, \beta_{s_{m}0} + \sum_{j=1}^p x_{ij} \beta_{s_{m}j}\Big) + \lambda [ \,  (1-\alpha ) \frac{1}{2}\sum_{j=1}^p \beta_{s_{m}j}^2 + \alpha \sum_{j=1}^p |\beta_{s_{m}j}| ],
\end{equation}
where \(\alpha\) controls the elastic net penalty and bridges the gap between LASSO (when \(\alpha =1\)) and ridge regression (when \(\alpha=0\)). If explanatory variables are correlated in groups, an \(\alpha\) around \(0.5\) tends to select the groups in or out altogether. For penalized least squares with elastic net penalty,
\[
\mathbf{\boldsymbol{\widehat{\beta}}}_{s_{m}} = \underset{\mathbf{\boldsymbol{\beta}}_{s_{m}}}{\operatorname{arg\,min}} \ \frac{1}{2n} \sum_{i=1}^n \Big(y_i - \beta_{s_{m}0} - \sum_{j=1}^p x_{ij} \beta_{s_{m}j}\Big)^2 + \lambda [ \,  (1-\alpha ) \frac{1}{2}\sum_{j=1}^p \beta_{s_{m}j}^2 + \alpha \sum_{j=1}^p |\beta_{s_{m}j}| ]\,
\]
This problem can also be solved using the coordinate descent algorithm. We omit the detailed solution which is a similar process to solve LASSO using the coordinate descent algorithm. However, it is worth mentioning that the expression for \(\widehat{\beta}_{s_{m}k}\) is as follows:
\[
\widehat{\beta}_{s_{m}k} = \frac {1}{1+\lambda(1-\alpha)} \left(\frac{\sum_{i=1}^n |\tilde{y_i}x_{ik}| }{\sum_{i=1}^n x_{ik}^2} - \frac{\lambda \alpha }{\frac{1}{n}\sum_{i=1}^n x_{ik}^2}\right)^+ \text{sgn}\Big( \sum_{i=1}^n \tilde{y_i}x_{ik} \Big)
\]

The mathematical foundation for the elastic net regularization can be found in \citet{de2009elastic}. It should also be noted that all the regularization schemes mentioned previously have simple Bayesian interpretations. See \citet{hastie2009} and \citet{james2013}. A review paper for statistical learning applications in general insurance can be found in \citet{parodi2012computational}.

Finally, we conclude that the hybrid tree-based models can be expressed in the following functional form:
\begin{equation} \label{eq:8}
H(\textbf{X},T(\Theta),\mathbf{\boldsymbol{\beta}_{s}} ) = 
\sum_{m=1}^{M} \underbrace{\beta_{f_{m}}}_\text{frequency} \textbf{X} \underbrace{\mathbf{\boldsymbol{\beta}}_{s_{m}} }_\text{severity} \underbrace{\mathbf{1}_{R_{m}}(\textbf{X})}_\text{risk class}
\end{equation}
From equation (\ref{eq:8}), we see that hybrid tree-based models can also be viewed as piecewise linear regression models. The tree-based algorithms divide the dataset into subsamples (or risk class), and linear regression models are then applied on these subsamples.

Algorithm \ref{alg:ht} summarizes the details of implementing hybrid tree-based models based on the \texttt{rpart} and \texttt{glmnetUtils} packages in R.

\begin{algorithm}[H]
\KwIn{Formula for the tree, formula for the linear model, training dataset $\textbf{x}$, cost-complexity parameter $cp$, maximum depth of final tree $maxdepth$, the threshold for the proportion of zeros in the node to be considered to build a linear model $zeroTreshold$, the choice of elastic net regularization $glmWhich$, penalty size $glmLambda$.}
\KwOut{Tree structure, linear models at each node, node information, $glmWhich$, $glmLambda$, fitted dataset.}
Grow a tree on a training dataset using recursive binary splitting. Use the stopping criterion $maxdepth$\;
Prune the tree using cost-complexity pruning with $cp$\;
Assign node information to each observation\;
\For{each terminal node}{
If zero claim exceeds the $zeroTreshold$ in the node, then we assign zero as the prediction\;
Else, we build a linear model using observations in this node. If the number of observations in this node is smaller than 40, then we assign the average of the response variable as the prediction; otherwise, we fit ordinary least squares with elastic net regularization model\;
}
Return Tree structure, linear models at each node, node information, $glmWhich$, $glmLambda$, fitted dataset\;
\caption{Implementation of the hybrid tree-based models}\label{alg:ht}
\end{algorithm}

\section{Simulation study} \label{sec:sim}

In this section, we conduct a simulation study to assess the performance of a simple hybrid tree-based model proposed in this paper relative to that of the Tweedie GLM. The synthetic dataset has been optimally created to replicate a real dataset. This synthetic dataset contains continuous explanatory variables and categorical explanatory variables with a continuous response variable sampled from Tweedie-like distribution. In some sense, the observed values of the response variable behave like a Tweedie, but some noise were introduced to the data for a fair comparison.

We begin with the continuous explanatory variables by generating sample observations from the multivariate normal with covariance structure
\[
\textbf{x} \sim N_p(0,\Sigma),
\]
where \(Cov(x)_{ij} = (0.5)^{i-j}\). Continuous explanatory variables close to each other has higher correlations, and the correlation effect diminishes with increasing distance between two explanatory variables. In real datasets, we can easily observe correlated explanatory variables among hundreds of possible features. It is also very challenging to detect the correlation between explanatory variables when an insurance company merges with third party datasets. Multicollinearity reduces the precision of the estimated coefficients of correlated explanatory variables. It can be problematic to use GLM without handling multicollinearity in the real dataset.

Categorical explanatory variables are generated by random sampling from the set of integers \((-3,-2,1,4)\) with equal probabilities. Categorical explanatory variables do not have any correlation structure among them.

We create \(10,000\) observations (with \(94.31\)\% zero claims) with \(60\) explanatory variables, including \(10\) continuous explanatory variables with relatively larger coefficients, \(10\) continuous explanatory variables with relatively smaller coefficients, \(10\) continuous explanatory variables with coefficients of \(0\), which means no relevance, \(10\) categorical explanatory variables with relatively larger coefficients, \(10\) categorical explanatory variables with relatively smaller coefficients, and \(10\) categorical explanatory variables with coefficients \(0\), which similarly implies no relevance. In effect, there are three groups of explanatory variables according to the size of the linear coefficients. In other words, there are strong signal explanatory variables, weak signal explanatory variables, and noise explanatory variables. Here are the true linear coefficients \(\boldsymbol{\beta_{Poisson}}\) and \(\boldsymbol{\beta_{Gamma}}\) used:
\[
\boldsymbol{\beta_{Poisson}}=(-0.1,\underbrace{0.5,\dots,0.5}_\text{10 continuous},\underbrace{0.1,\dots,0.1}_\text{10 continuous},\underbrace{0,\dots,0}_\text{10 continuous},\underbrace{-0.5,\dots,-0.5}_\text{10 categorical},\underbrace{0.1,\dots,0.1}_\text{10 categorical},\underbrace{0,\dots,0}_\text{10 categorical})^T
\]

\[
\boldsymbol{\beta_{Gamma}}=(6,\underbrace{0.5,\dots,0.5}_\text{10 continuous},\underbrace{-0.1,\dots,-0.1}_\text{10 continuous},\underbrace{0,\dots,0}_\text{10 continuous},\underbrace{0.5,\dots,0.5}_\text{10 categorical},\underbrace{-0.1,\dots,-0.1}_\text{10 categorical},\underbrace{0,\dots,0}_\text{10 categorical})^T
\]
The first coefficients refer to the intercepts. Because of the equivalence of the compound Poisson-gamma with Tweedie distributions, as discussed in Section (\ref{sec:intro}), we generated samples drawn from Poisson for frequency and gamma for severity.

To illustrate possible drawbacks using Tweedie GLM, we use different linear coefficients for the frequency part and the severity part. The absolute value of the linear coefficients is the same but with a different sign. In real-life, there is no guarantee that the explanatory variables have equal coefficients in both frequency and severity parts. The response variable generated by the compound Poisson-gamma distribution framework using modified \texttt{rtweedie} function in \texttt{tweedie} R package with frequency and severity as components is presented as follows:

\noindent \textbf{Frequency part}: Poisson distribution with a log link function:
\[
N_i\sim Poisson(\lambda) \text{ with } \lambda = \lambda_i \text{ for given } x_i,
\]
where $\displaystyle \lambda_i=\frac{\exp(\textbf{x}_i\boldsymbol{\beta_{Poisson}})^{2-\text{power}}}{\phi \times (2-\text{power})}$ , with  $\text{power}=1.5$ and $\phi=2$.

\medskip

\noindent \textbf{Severity part}: Gamma distribution with a log link function:
\[
Y_j\sim Gamma(\alpha,\beta)  \text{ with }\alpha = \frac{2 - \text{power}}{\text{power} - 1}  \text{ and }\beta=\beta_i  \text{ for given }x_i,
\]
where $\displaystyle \beta_i=\frac{\exp(\textbf{x}_i\boldsymbol{\beta_{Gamma}})^{1-\text{power}}}{\phi \times (\text{power}-1)}$.

\medskip

\noindent \textbf{Response variable}:
\[
Y^r_i = \sum_{j=1}^{N_i}Y_j\text{ for given } x_i, \quad \text{where} \ j=1, \dots , N_i,
\]
where the superscript $r$ has been used to distinguish the response and the gamma variables.

We introduce noise to the true Tweedie model in the following ways: multicollinearity among continuous variables, zero coefficients for non-relevant explanatory variables, different coefficients for frequency and severity part, and added white noise to positive claims. These noises are what make the real-life data deviate from the true model in practice. It makes for an impartial dataset that can provide for a fair comparison.

Once we have produced the simulated (or synthetic) dataset, we apply both Tweedie GLM and hybrid tree-based models to predict values of the response variables, given the available set of explanatory variables. The prediction performance between the two models is then compared using several validation measures as summarized in Table \ref{tab:SM}. In Appendix A, Table \ref{tab:VM} describes details for validation measures.

\begin{table}[!htbp]
\begin{center}
\caption{Prediction accuracy using the simulation dataset.} \label{tab:SM}
\resizebox{!}{!}{
\begin{tabular}{lrrrrr}
\toprule
Validation measure & $GINI$ & $R^2$ & $CCC$ & $RMSE$ & $MAE$  \\ 
\midrule
Tweedie GLM             & 0.95  & 0.16 & 0.47  & 0.89 & 0.07  \\ 
Hybrid Tree-based Model & 0.97  & 0.29 & 0.36  & 0.81 & 0.06 \\
\bottomrule
\end{tabular}}
\end{center}
\end{table}

From Table \ref{tab:SM}, we find that the Tweedie GLM performs worse even under the true model assumption, albeit the presence of noise. However, hybrid tree-based models with elastic net regression are able to automatically perform coefficient shrinkage via L2 norm and variable selection based on the tree-structure algorithm and L1 norm. Hybrid tree-based models provide more promising prediction results than the now popular Tweedie GLM. Even under the linear assumption as in our simulation dataset, hybrid tree-based models still perform better.

\section{Empirical application} \label{sec:APP}

We compare Tweedie GLM with hybrid tree-based models using one coverage group from the LGPIF data. This coverage group is called building and contents (BC) which provides insurance for buildings and the properties for local government entities including counties, cities, towns, villages, school districts, and other miscellaneous entities. We draw observations from years 2006 to 2010 as training dataset and year 2011 as test dataset. Table \ref{tab:1} and Table \ref{tab:2} summarize training dataset and test dataset respectively.

\begin{table}[!htbp]
\begin{center}
\caption{Summary statistics of the variables for the training dataset, 2006-2010.} \label{tab:1}
\resizebox{\linewidth}{!}{
\begin{tabular}{llrrrrrr}
\toprule

Response \\   
variables & Description & Min. & 1st Q & Mean & Median & 3rd Q & Max. \\ 

\midrule

ClaimBC & BC claim amount in million.
& $0$
& $0$
& $0.02$
& $0$
& $0$
& $12.92$ \\ 

\midrule

Continuous \\   
variables &  &  &  &  &  &  &  \\ 

\midrule

CoverageBC & Log of BC coverage amount.
& $-7.95$
& $0.78$ 
& $2.12$
& $2.42$ 
& $3.6$ 
& $7.8$\\

lnDeductBC & Log of BC deductible level.
& $6.21$ 
& $6.21$ 
& $7.15$
& $6.91$ 
& $7.82$ 
& $11.51$\\

\midrule

Categorical \\
variables  &  \multicolumn{5}{l}{}  & \multicolumn{2}{r}{Proportions} \\

\midrule
                        
NoClaimCreditBC    & \multicolumn{5}{l}{Indicator for no BC claims in previous year.} 
& \multicolumn{2}{r}{32.93 \%} \\

TypeCity    & \multicolumn{5}{l}{EntityType: City.} 
& \multicolumn{2}{r}{14.03 \%} \\

TypeCounty   & \multicolumn{5}{l}{EntityType: County.}            
& \multicolumn{2}{r}{5.80 \%} \\

TypeMisc   & \multicolumn{5}{l}{EntityType: Miscellaneous.}    
& \multicolumn{2}{r}{10.81 \%} \\

TypeSchool   & \multicolumn{5}{l}{EntityType: School.}      
& \multicolumn{2}{r}{28.25 \%} \\

TypeTown   & \multicolumn{5}{l}{EntityType: Town.}
& \multicolumn{2}{r}{17.33 \%} \\

TypeVillage   & \multicolumn{5}{l}{EntityType: Village.}
& \multicolumn{2}{r}{23.78 \%} \\

\bottomrule

\end{tabular}}
\end{center}
\end{table}

\begin{table}[!htbp]
\begin{center}
\caption{Summary statistics of the variables for the test dataset, 2011.}
\label{tab:2}
\resizebox{\linewidth}{!}{
\begin{tabular}{llrrrrrr}
\toprule

Response \\   
variables & Description & Min. & 1st Q & Mean & Median & 3rd Q & Max. \\ 

\midrule

ClaimBC & BC claim amount in million.
& $0$
& $0$
& $0.02$
& $0$
& $0$
& $2.75$ \\ 

\midrule

Continuous \\   
variables &  &  &  &  &  &  &  \\ 

\midrule

CoverageBC & Log of BC coverage amount.
& $-7.84$
& $0.91$ 
& $2.27$
& $2.56$ 
& $3.74$ 
& $7.78$\\

lnDeductBC & Log of BC deductible level.
& $6.21$ 
& $6.21$ 
& $7.24$
& $6.91$ 
& $7.82$ 
& $11.51$\\

\midrule

Categorical \\
variables  &  \multicolumn{5}{l}{}  & \multicolumn{2}{r}{Proportions} \\

\midrule
                        
NoClaimCreditBC    & \multicolumn{5}{l}{Indicator for no BC claims in previous year.} 
& \multicolumn{2}{r}{50.87 \%} \\

TypeCity    & \multicolumn{5}{l}{EntityType: City.} 
& \multicolumn{2}{r}{14.06 \%} \\

TypeCounty   & \multicolumn{5}{l}{EntityType: County.}            
& \multicolumn{2}{r}{6.48 \%} \\

TypeMisc   & \multicolumn{5}{l}{EntityType: Miscellaneous.}    
& \multicolumn{2}{r}{11.42 \%} \\

TypeSchool   & \multicolumn{5}{l}{EntityType: School.}      
& \multicolumn{2}{r}{27.67 \%} \\

TypeTown   & \multicolumn{5}{l}{EntityType: Town.}
& \multicolumn{2}{r}{16.53 \%} \\

TypeVillage   & \multicolumn{5}{l}{EntityType: Village.}
& \multicolumn{2}{r}{23.84 \%} \\

\bottomrule

\end{tabular}}
\end{center}
\end{table}

\subsection{Tweedie GLM} \label{tweedie-glm}

We replicate the Tweedie GLM as noted in \citet{frees2016mv}. The linear coefficients for the Tweedie GLM has been provided in Table A11 in \citet{frees2016mv}. It was pointed out that Tweedie GLM may not be ideal for the dataset after visualizing the cumulative density function plot of jittered aggregate losses as depicted in Figure A6 of \citet{frees2016mv}.

\subsection{CART algorithm} \label{cart-algorithm}

We perform grid search on the training dataset with 10-fold cross-validation. The final model has hyperparameters set with 8 minimum number of observations in a region for the recursive binary split to be attempted and cost-complexity pruning parameter (\(cp\)) is 0.05.

\subsection{Calibration results of hybrid tree-based models} \label{hybrid}

We use two hybrid tree-based models for this application. For the binary classification, we use the CART algorithm, and at the terminal node, we use either GLM with Gaussian family or elastic net regression with Gaussian family. Similar to the CART part, we perform a grid search with 10-fold cross-validation to find the optimal set of hyperparameter values. Hybrid tree-based models with simple GLM with Gaussian family (HTGlm) has the following hyperparameter setting: \(cp\) is \(0.0001\), maximum depth of the tree (\(maxdepth\)) is \(8\), and the threshold for the percentage of the zeros in the node to determine whether to build the linear model or assign \(zero\) as the prediction (\(zeroThreshold\)) is \(0.25\). Hybrid tree-based models with elastic net regression (HTGlmnet) has the following hyperparameter setting: \(cp\) is \(0.0002\), \(maxdepth\) is \(10\), \(zeroThreshold\) is \(0.23\), the balance between L1 norm and L2 norm (\(glmWhich\)) is 1, and size of the regularization parameter by the best cross-validation result (\(glmLambda\)) is \(lambda.min\). These hyperparameter settings are chosen to have optimal RMSE in the model calibration.

\subsection{Model performance} \label{model-performance}

We use several validation measures to examine the performance of different models just as what was done in the simulation. To make for an easier comparison, we added the heat map in addition to the prediction accuracy table based on various validation measures. We present results separately for the training and test datasets. Rescaled value of 100 (colored blue) indicates it is the best model among the comparisons; on the other hand, rescaled value of 0 (colored red) indicates it is the worst one. For further details on such heat map for validation, see \citet{quan2018predictive}.

\smallskip

\begin{figure}[htbp]
    \centering
    \begin{subfigure}{0.8\textwidth}
      \includegraphics[width=\linewidth]{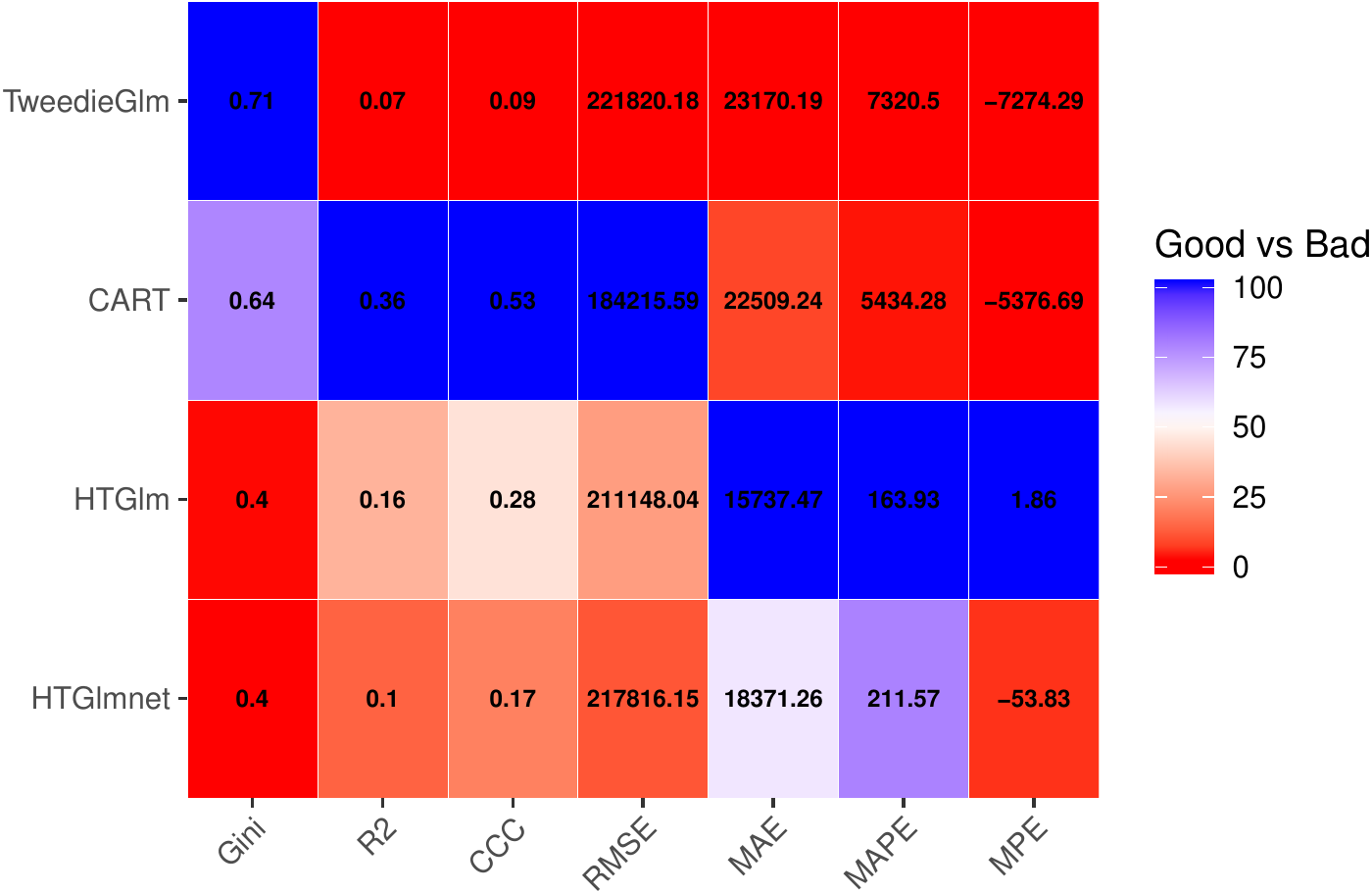}
      \caption{Model performance based on training dataset}
      \label{fig:trainHeat}
    \end{subfigure}
    \begin{subfigure}{0.8\textwidth}
      \includegraphics[width=\linewidth]{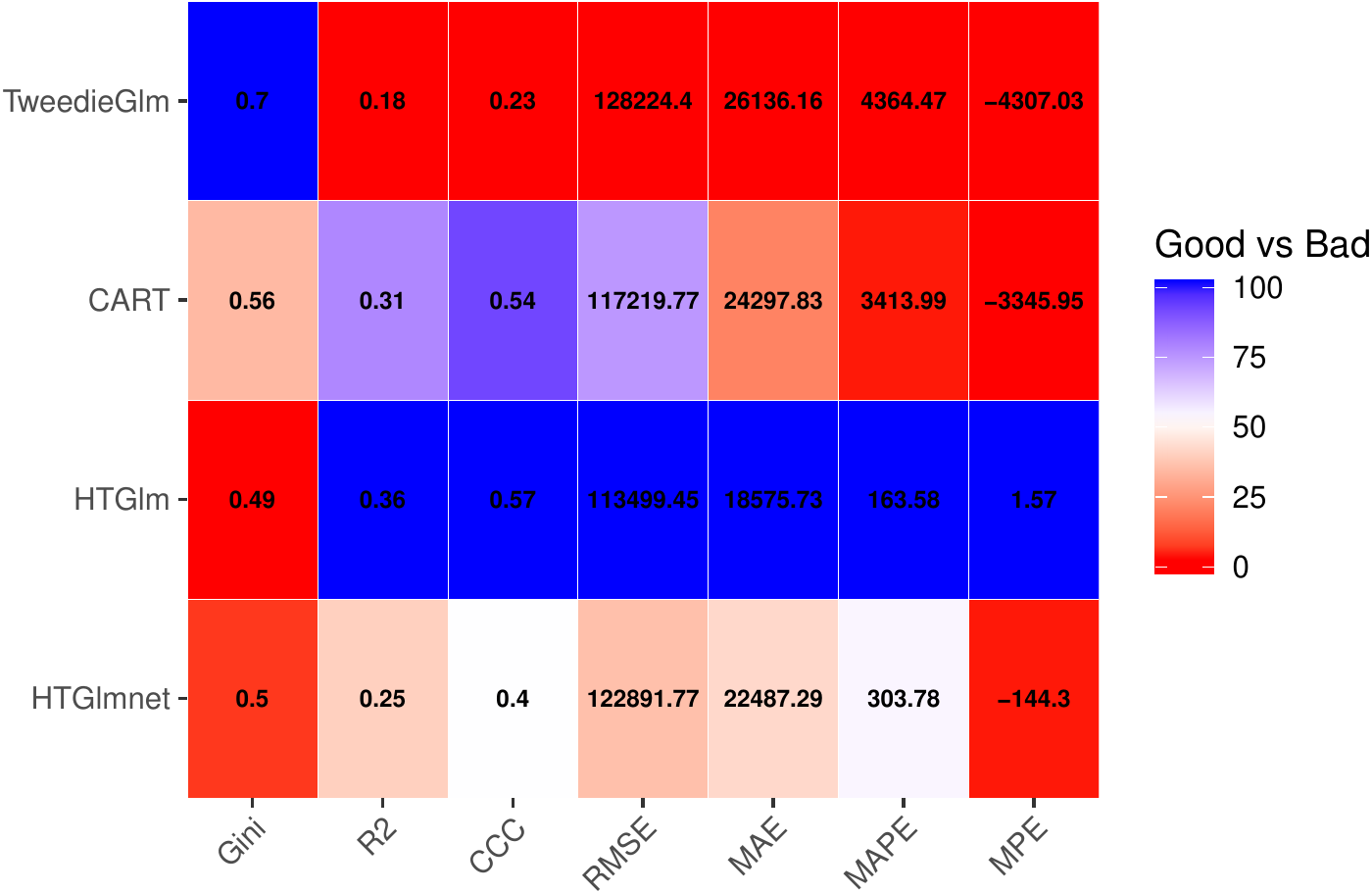}
      \caption{Model performance based on test dataset}
      \label{fig:testHeat}
    \end{subfigure}
   \caption{A heat map of model comparison based on various validation measures.}
 \label{fig:vm}
\end{figure}

From Figure \ref{fig:trainHeat}, we find that Tweedie GLM has the worst performance of model fit in general, and the CART algorithm and HTGlm has better performance on the model fit. From Figure \ref{fig:testHeat}, we find that HTGlm has the best performance on the test dataset. It also shows that the hybrid tree-based models prevents overfitting compared to traditional regression trees using the CART algorithm. Hybrid tree-based models perform consistently well for both training dataset and test dataset.  Because hybrid tree-based models are considered piecewise linear models, the consistency is inherited from the linear model. We can also note  that HTGlmnet has worse performance when compared to HTGlm, for both the training and test datasets. The advantage of regularization does not demonstrate significant effects on the LGPIF dataset because of the limited number of explanatory variables. However, in the simulation study, we showed the advantage of the regularization because the data has a more significant number of explanatory variables. The simulated data may be a better replication of real-life situation. In summary, hybrid tree-based models outperform Tweedie GLMs.

\section{Visualization and interpretation} \label{sec:visual}

\begin{figure}[htbp]
\centering
\includegraphics[width=\linewidth]{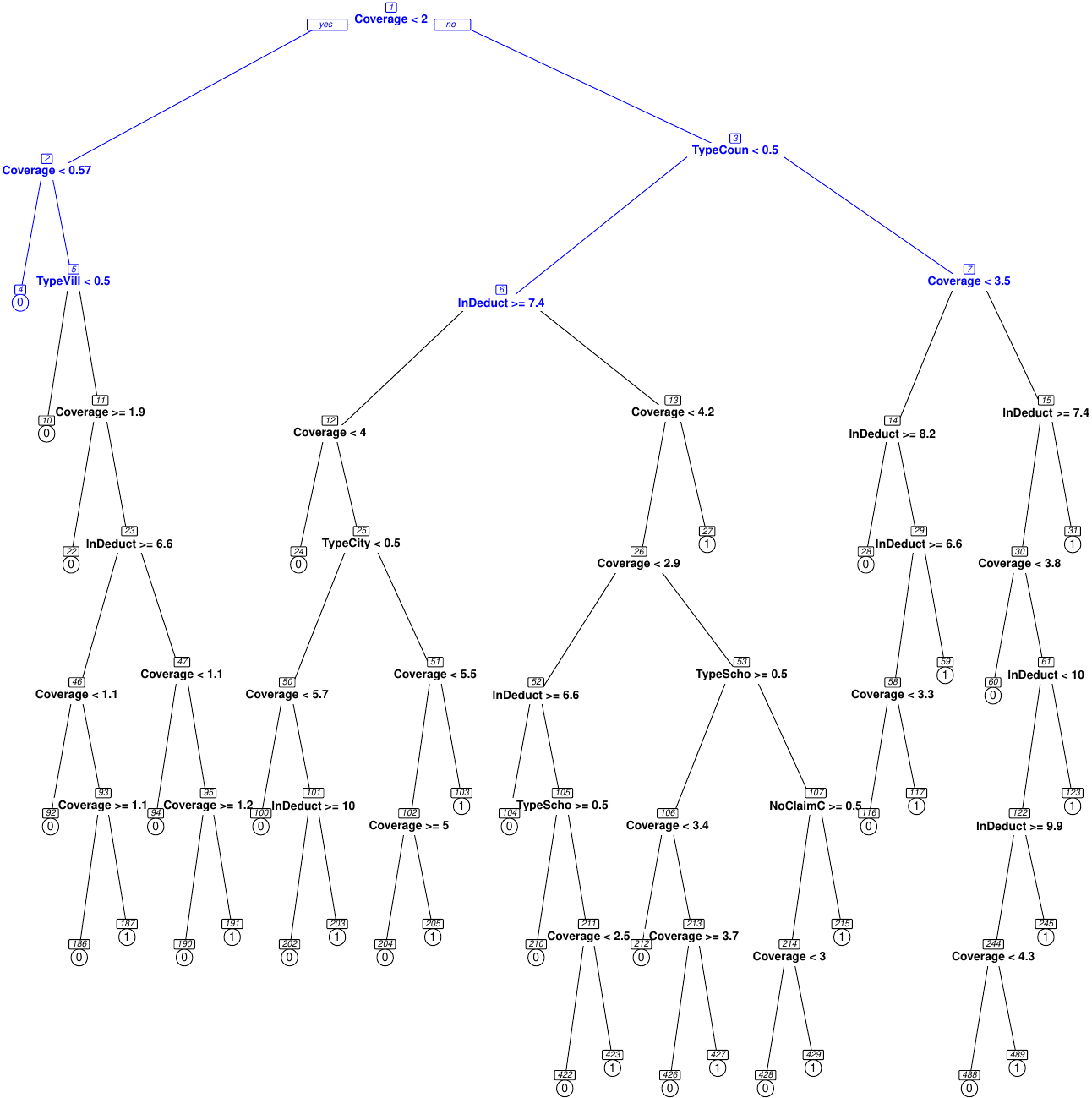}
\caption{Tree paths with highlighted nodes.} \label{fig:treePath}
\end{figure}

\begin{figure}[htbp]
\centering
\includegraphics[width=\linewidth]{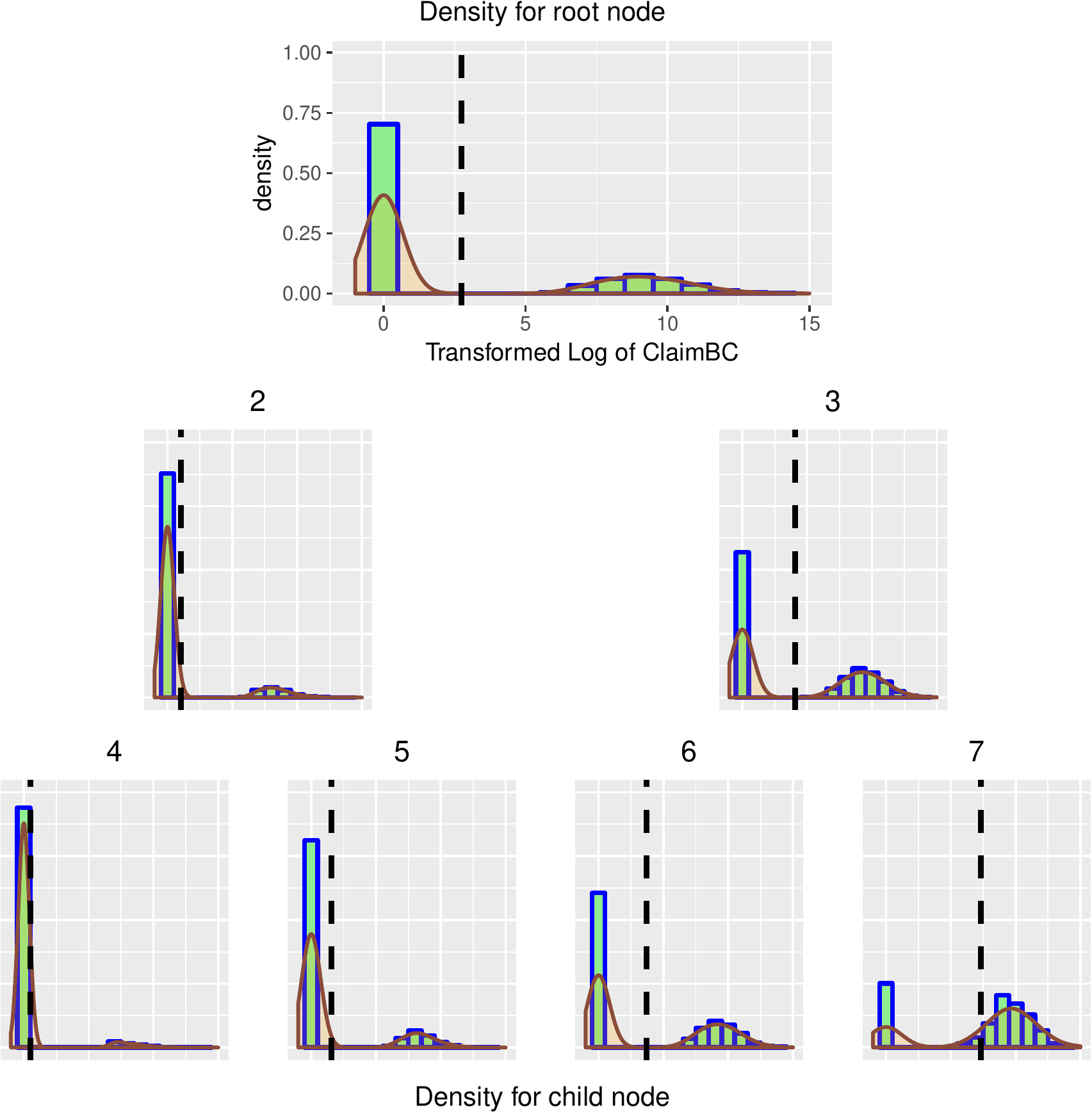}
\caption{Classification tree for the frequency.} \label{fig:density}
\end{figure}

The frequency component of hybrid tree-based models can clearly be visualized as tree structure like a simple CART. Here, we use HTGlm to illustrate the visualization and interpretation of hybrid tree-based models. In Appendix B, Figure \ref{fig:tree} shows the classification tree model for the frequency. We can follow the tree path and construct the density plot for the response variable for each node. Figure \ref{fig:treePath} shows the classification tree and we highlighted (blue) nodes two depths away from the root. Consequently, we have seven nodes including the root nodes. Here we denote the node with numbers for the purpose of identification. Node four (4) is the terminal node and others are intermediate nodes. Figure \ref{fig:density} shows the density of the log of the response variable for the seven nodes mentioned in Figure \ref{fig:treePath}. The root node has zero point mass as expected and the black dashed line is the mean of the response variable at the node. After the first split using CoverageBC, the mean shifted towards a different direction and the percentage of the point mass at zero is also changed in a different direction. At the fourth terminal node, we can barely see any positive claims with the mean shifting towards zero and this terminal node ultimately assigns zero as the prediction. We can see the classification tree algorithm, done by recursive binary splitting, divides the dataset into subspaces that contain more similar observations. After dividing the space of explanatory variables into subspaces, it may be more appropriate to make the distribution assumptions and apply the linear model to the subsamples. Hybrid tree-based models are, in some sense, employ an algorithm that `divide and conquer' and thus it can be a promising solution for imbalance datasets.

\smallskip

\smallskip

\smallskip

\begin{figure}[htbp]
\centering
\includegraphics{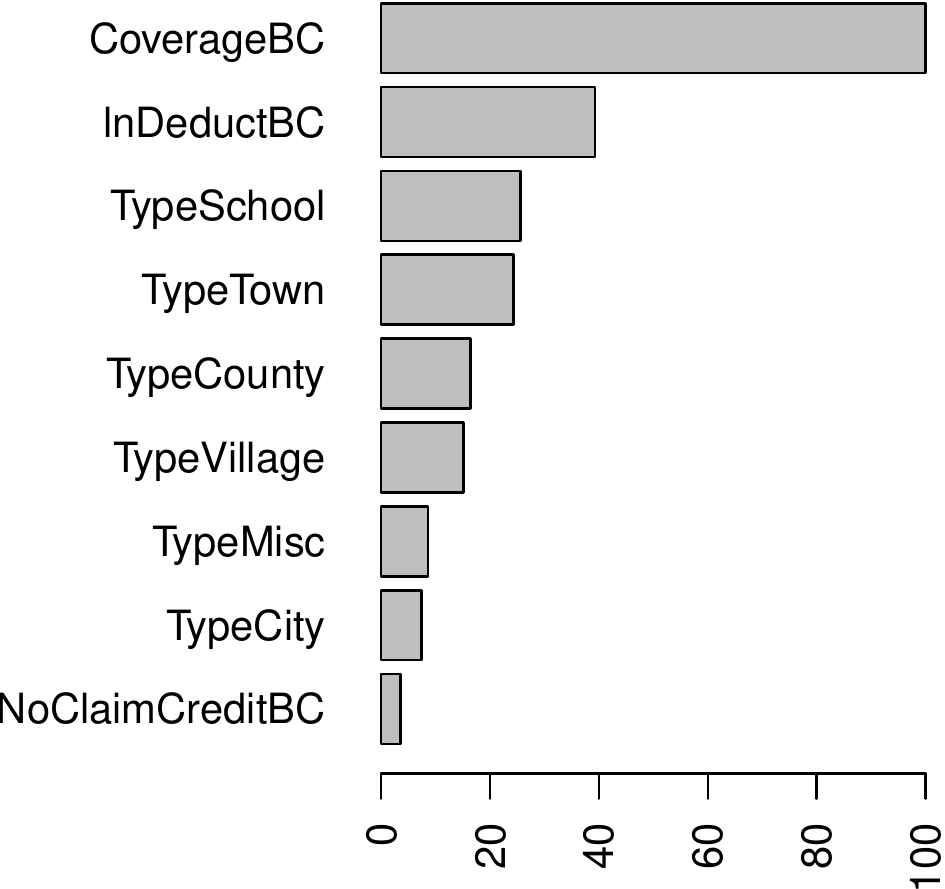}
\caption{Variable importance for the claim frequency.} \label{fig:VI}
\end{figure}

Figure \ref{fig:treePath} shows the path to terminal nodes and we can see a zero or a one, where a linear model is then fitted, for each terminal node. The tree structure and path can provide an indication of the risk factors based on the split of the explanatory variables. Figure \ref{fig:VI} shows variable importance calculated from the classification tree model for the frequency part. Coverage and deductible information are much more important than location and previous claim history to identify whether claim happened or not. At each terminal node, we can also extract the regression coefficients and interpret the estimated linear regression model. Table \ref{tab:beta} provides the regression coefficients at each of the non-zero terminal nodes. The node number has the same label as in Figure \ref{fig:treePath}. For example, the terminal node 245 can be reached as follows: \(CoverageBC\geq2\), \(TypeCounty\), \(Coverage\geq3.5\), \(lnDeductBC\geq7.4\), \(CoverageBC\geq3.8\), \(lnDeductBC<10\), \(lnDeductBC<9.9\), and we estimated the linear model with regression coefficients: \(Intercept=-172854\), \(CoverageBC=15251\), \(lnDeductBC=16777\), \(NoClaimCreditBC=-16254\). At the terminal node, the linear model is built that solely relies on the prediction accuracy from cross-validation and based on the regularized algorithm. It does not rely on statistical inference, and indeed, some regression coefficients are not statistically significant. However, we can tune hybrid tree-based models by balancing the prediction accuracy and statistical inference.

\begin{table}[!htbp]
\begin{center}
\caption{Regression coefficients at the terminal nodes.} \label{tab:beta}
\resizebox{\linewidth}{!}{
\begin{tabular}{lrrrrrrr}
\toprule

Terminal node & 245 & 31 & 123 & 27 & 203 & 103 & 187 \\ 
 & \multicolumn{7}{c}{Estimates} \\ 

\midrule

(Intercept)     & -172854 & 24263  & 7922014  & 150264 & -1907177 & 155878 & -303658 \\ 
TypeCity        &         &        &          & 17218  &          &        &         \\
TypeCounty      &         &        &          &        &          &        &         \\
TypeMisc        &         &        &          &        & 62714    &        &         \\
TypeSchool      &         &        &          &        &          &        &         \\
TypeTown        &         &        &          &        &          &        &         \\
CoverageBC      & 15251   & 26601  & 1102629  & 9454   & 447654   & 2772   & 287021  \\
lnDeductBC      & 16777   & -15412 & -1232514 & -25246 & -70542   & -1526  &         \\
NoClaimCreditBC & -16254  & -34160 &          & -21825 & -67030   & -13229 & -13632  \\

\bottomrule

\end{tabular}}
\end{center}
\end{table}

\section{Conclusions} \label{sec:conclude}

With the development of data science and increased computational capacity, insurance companies can benefit from existing large datasets collected over a stretched period of time. On the other hand, it is also challenging to build a model on ``big data.'' It is especially true for modeling claims of short-term insurance contracts since the data set presents many unique characteristics. For example, claim datasets typically have a large proportion of zero claims that lead to imbalances in classification problems, a large number of explanatory variables with high correlations among continuous variables, and many missing values in the datasets. It can be problematic if we directly use traditional approaches like Tweedie GLM without additional modifications on these characteristics. Hence, it is often not surprising to find a less desirable prediction accuracy of conventional methods for real-life datasets.

In this paper, to fully capture the information available in large insurance datasets, we propose the use of hybrid tree-based models as a learning machinery and an expansion to the available set of actuarial toolkits for claims prediction. We have reiterated the many benefits that can be derived from such a hybrid structure. The use of classification trees to model the frequency component has the advantages of being able to handle imbalanced classes and to naturally subdivide portfolio into risk classes, as well as the many benefits of tree-based models. See also \cite{quan2018predictive}. The use of a regression model for the severity component provides the flexibility of identifying factors that are important predictors for accuracy and interpretation. Finally, the hybrid specification captures the benefits of tuning hyperparameters at each step of the algorithm. The tuning parameters can be adjusted not only for prediction accuracy but also for possibly reaching for specific business objectives in the decision making process. The transparency and interpretability are advantages to convince interested parties, such as investors and regulators, that hybrid tree-based models are much more than just black box.

We examine the prediction performance of our proposed models using datasets produced from a simulation study and real-life insurance portfolio. We focused on comparison to the Tweedie GLM since this has become widespread for modeling claims arising from short-term insurance policies. Broadly speaking, hybrid-tree models can outperform such Tweedie GLM without loss of interpretability. However, there are opportunities to improve and to widen the scope of this hybrid tree structure that can be explored for further research.

\section*{Acknowledgment}

We thank the financial support of the Society of Actuaries through its Centers of Actuarial Excellence (CAE) grant for our research project on \textit{Applying Data Mining Techniques in Actuarial Science}.

\smallskip

\appendix
\section*{Appendix A. Performance validation measures}

The following table provides details of the various validation measures that were used throughout this paper to compare different predictive models.

\begin{table}[!htbp]
\caption{Performance validation measures.} \label{appendix-a}
\vspace*{5mm}
\centering
\label{tab:VM}
\resizebox{\linewidth}{!}{
\begin{tabular}{lll}
\toprule
Validation measure & Definition & Interpretation \\
\midrule
Gini Index    & $Gini = 1 - \dfrac{2}{N-1} \left(N - \dfrac{\sum_{i=1}^N \; i\tilde{y}_i}{\sum_{i=1}^N \tilde{y}_i}\right)$  &  Higher $Gini$ is better. \\
 & where $\tilde{y}$ is the corresponding to $y$ after & \\
 & ranking the corresponding predicted values $\widehat{y}$.  & \\
\midrule
Coefficient of Determination    & $R^2 = 1 - \dfrac{\sum_{i=1}^N (\widehat{y}_i - y_i)^2}{\sum_{i=1}^N \left(y_i - \dfrac{1}{n} \sum_{i=1}^n y_i\right)^2}$ & Higher $R^2$ is better. \\
 & where $\widehat{y}$ is predicted values. & \\
\midrule
Concordance Correlation   &  $CCC = \displaystyle \frac{2 \rho \sigma_{\widehat{y}_{i}} \sigma_{y_{i}}}{\sigma_{\widehat{y}_{i}}^2 + \sigma_{y_{i}}^2 + (\mu_{\widehat{y}_{i}} - \mu_{y_{i}})^2}$ 
 &  Higher $CCC$ is better. \\
 Coefficient & where $\mu_{\widehat{y}_{i}}$ and $\mu_{y_{i}}$ are the means & \\
 & $\sigma^2_{\widehat{y}_{i}}$ and $\sigma^2_{y_{i}}$ are the variances & \\
 & $\rho$ is the correlation coefficient & \\
\midrule
Root Mean Squared Error    & $RMSE = \sqrt{\dfrac{1}{N} \sum_{i=1}^N  (\widehat{y}_{i}-{y}_{i})^2 }$ & Lower $RMSE$ is better \\
\midrule
Mean Absolute Error    & $MAE = \dfrac{1}{N} \sum_{i=1}^N  \left| \widehat{y}_{i}-{y}_{i} \right|$ & Lower $MAE$ is better. \\
\midrule
Mean Absolute Percentage Error  & $MAPE = \dfrac{1}{N} \sum_{i=1}^N  \left| \dfrac{\widehat{y}_{i}-{y}_{i}}{y_{i}} \right|$ & Lower $MAPE$ is better. \\
\midrule
Mean Percentage Error    & $MPE = \dfrac{1}{N} \sum_{i=1}^N   \dfrac{\widehat{y}_{i}-{y}_{i}}{y_{i}}$ & Lower $\vert MPE\vert$ is better. \\
\bottomrule
\end{tabular}}
\end{table}

\newpage

\section*{Appendix B. Classification tree for the claim frequency} \label{appendix-b}

\begin{figure}[htbp]
\centering
\includegraphics{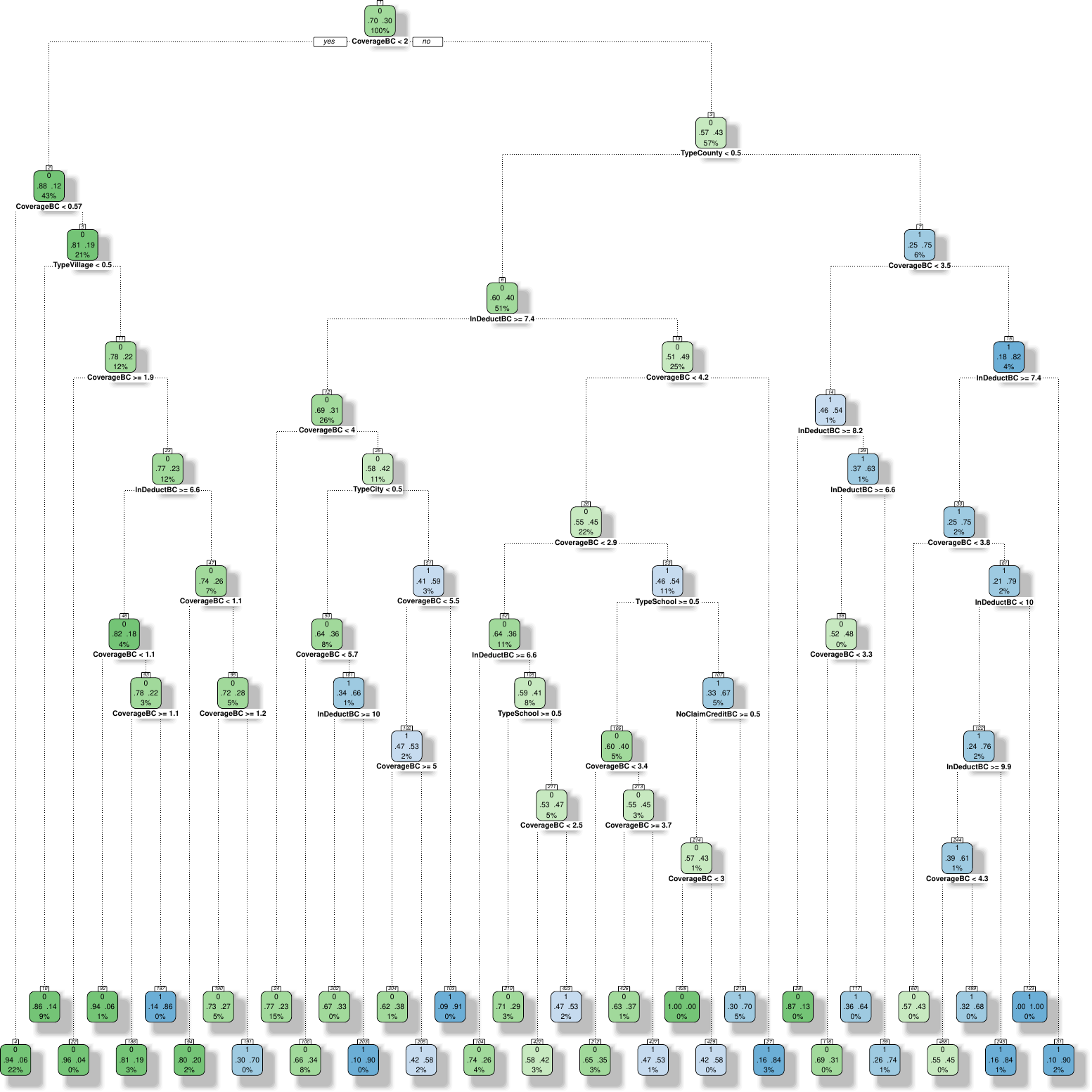}
\caption{Classification tree for the claim frequency.} \label{fig:tree}
\end{figure}

\newpage

\bibliographystyle{apalike}
\bibliography{hybridtree.bib}

\end{document}